\documentclass{JHEP3}

\newcommand{\3}{\mbox{${\bf \underline{3}}$}}
\newcommand{\s}{\mbox{${\bf \underline{1}}$}}
\newcommand{\spr}{\mbox{${\bf \underline{1}'}$}}
\newcommand{\sppr}{\mbox{${\bf {\underline{1}''}}$}}

\title{$A_4$ flavour symmetry breaking scheme for understanding quark and neutrino mixing angles}

\author{Xiao-Gang He\\ NCTS/TPE, Department of Physics, National Taiwan University, Taipei,
 Taiwan, 10764, Republic of China\\
E-mail: \email{hexg@phys.ntu.edu.tw}}
\author{Yong-Yeon Keum\\ NCTS/TPE, Department of Physics, National Taiwan University, Taipei,
Taiwan, 10764, Republic of China\\
E-mail: \email{yykeum@phys.ntu.edu.tw}}
\author{Raymond R. Volkas\\ School of Physics, Research Centre for High Energy Physics, The University
of Melbourne, Victoria 3010, Australia\\ E-mail:
\email{r.volkas@physics.unimelb.edu.au}}

\abstract{We propose a spontaneous $A_4$ flavour symmetry
breaking scheme to understand the observed pattern of quark and
neutrino mixing.  The fermion mass eigenvalues are arbitrary, but
the mixing angles are constrained in such a way that the overall
patterns are explained while also leaving sufficient freedom to
fit the detailed features of the observed values, including
CP-violating phases.  The scheme realises the proposal of Low and
Volkas to generate zero quark mixing and tribimaximal neutrino
mixing at tree level, with deviations from both arising from small
corrections after spontaneous $A_4$ breaking.  In the neutrino
sector, the breaking is $A_4 \to Z_2$, while in the quark and
charged-lepton sectors it is $A_4 \to Z_3 \cong C_3$.  The full
theory has $A_4$ completely broken, but the two different unbroken
subgroups in the two sectors force the dominant mixing patterns to
be as stated above.  Radiative effects within each sector are
shown to deviate neutrino mixing from tribimaximal, while
maintaining zero quark mixing. Interactions between the two
sectors -- ``cross-talk'' -- induce nonzero quark mixing, and additional deviation from
tribimaximal neutrino mixing.  We discuss the vacuum alignment challenge
the scenario faces, and suggest three generic ways to approach the problem.
We follow up one of those ways by
sketching how an explicit model
realising the symmetry breaking structure may be constructed.}


\preprint{}

\begin{document}

\section{Introduction}

The explanation of flavour is one of the most profound goals in
the construction of standard model (SM) extensions.  There are
several aspects to the overall puzzle:  Why three families of
quarks and leptons? Why the specific mixing patterns observed? Can
we understand quark and lepton mass eigenvalues? Why are neutrinos
so light?  A priori, it is not clear if these aspects of the
flavour problem should be treated organically, or they can be
solved piecemeal.  In this paper, we show that the observed mixing
angle patterns suggest an underlying flavour symmetry breaking
structure based on the discrete group $A_4$, while also
incorporating the see-saw explanation for why neutrinos are
especially light.\footnote{$A_4$ is the alternating group of order
four, defined as the set of all even permutations of four objects.
Geometrically, it is the symmetry group of the tetrahedron. See
the appendix A for a list of the basic results we will use in this
paper, and Refs.\cite{a4,a4alterali,a4he,a4zee,a4ma} for further
discussion.}

It is interesting that these features can be
understood without needing to know why the fermions come in three
families, and without drawing any connection with an explanation
for the mass eigenvalues.  The latter will be arbitrary in our
scheme.  Note that in the neutrino sector, most of our current knowledge about
masses and mixing angles comes from neutrino oscillation data which
provide no direct information about the absolute values of
neutrino masses, while detailed information about neutrino mixing
has been obtained.  The analysis to be presented may provide crucial
information for understanding the mixing mechanism in both the
lepton and quark sectors.  The present neutrino
data can be accommodated fairly well \cite{data} by the so-called
tribimaximal mixing \cite{tribimaximal} matrix.  We shall therefore use
tribimaximal neutrino mixing as the lowest order approximation and
study allowed deviations within our $A_4$ structure.  In the quark
sector, we set up the mixing matrix to be the identity matrix at lowest
order, and then generate non-trivial mixing through corrections.

The main point of this paper is to argue for a certain
symmetry-breaking structure, rather than to advocate a specific
model realising it (we shall call this the ``dynamical completion''
problem).  Given our present ignorance, at the
experimentally-verified level, of the precise dynamics nature
chose to break the electroweak symmetry, we feel that it is of
considerable value to begin by studying flavour
symmetry breaking in a way as independent of dynamics as possible.
The resulting symmetry-based understanding may have more long-term
value than, for example, explicit Higgs-potential realisations.
Indeed, the insights so gained can be used to guide subsequent
model-building.  The latter may encompass conventional Higgs
models, as well as brane-world realisations and other schemes,
according to the skill and taste of the model-builder.

Nevertheless, we shall find it convenient to think in terms of
Higgs fields and their expectation values, and for the sake of completeness
we shall also briefly discuss the dynamical completion issue and suggest
possible solutions.

Our ideas were seeded by two considerations from the literature.
First, several authors have recently explored very interesting
connections between flavour $A_4$ \cite{a4alterali,a4he,a4zee,a4ma}
and the tribimaximal neutrino mixing matrix \cite{tribimaximal}.
Second, a conjecture has been proposed by Low and Volkas on the
relationship between quark and lepton mixing \cite{lowvolkas}.  It
is often remarked that these two sectors are jarringly different:
quark mixing reveals three small mixing angles, while lepton
mixing requires two large angles (one consistent with maximal),
and one small angle (consistent with zero).  The conjecture begins
by noting that each of the mixing matrices is of the form
$V_1^{\dagger} V_2$, where the $V_i$ are the left-diagonalisation
matrices of the species involved.  The proposal is that due to
symmetry, the $V_i$ matrices for up-quarks and down-quarks are
identical, leading to a trivial Cabibbo-Kobayashi-Maskawa (CKM)
matrix \cite{ckm}, but with each quark $V_i$ having large
off-diagonal entries.  Quark mixing is ``trying'' to be large, but
the effects are exactly cancelled due to a symmetry. In the lepton
sector, one then proposes that one of the diagonalisation matrices
-- the neutrino one in the original conjecture and subsequently in
the present paper -- takes a different form from the other three
due to different symmetry constraints.  There is no perfect
cancellation, and thus Maki-Nakagawa-Sakata-Pontecorvo
(MNSP) \cite{mnsp} mixing contains large angles.  To agree with
experiment, and to hold the promise of an underlying symmetry, the
tribimaximal form for the MNSP matrix is selected.  Deviations
from diagonal CKM and tribimaximal MNSP then arise from
corrections after symmetry breaking.  We show how this conjecture
can be realised in our $A_4$ scheme.

Our approach delivers some interesting quantitative relations
between mixing angles and CP violating phases, and relates the
neutrino mixing angle $\theta_{13}$ to other deviations from the
tribimaximal MNSP form.

In the next section we define the field content of the scheme, and
explain how the dominant tree-level mixing matrices arise.
Section 3 then explains how a class of radiative corrections --
those intrinsic to the neutrino sector on its own, and the
charged-fermion sector on its own -- alters the tree-level
picture: the tribimaximal pattern is modified, but CKM mixing is
still absent. In Sec.\ 4, interactions between the sectors are
used to generate CKM mixing and, generically, to induce additional
deviation for tribimaximal neutrino mixing.  Section 5
discusses the dynamical completion challenge,
while Sec.\ 6 is a conclusion.  Appendix A states the basic
$A_4$ results we shall use, App.\ B lists the Higgs potential of the
minimal model, while App.\ C discusses a supersymmetric dynamical
completion.

\section{The scheme and tree-level results}

The symmetry group of our scheme is $G \otimes X$, where
\begin{equation}
G = SU(3)_c \otimes SU(2)_L \otimes U(1)_Y \otimes A_4,
\end{equation}
and the usual SM gauge group is augmented by an $A_4$ flavour
symmetry plus an auxilliary symmetry $X$ whose nature and role shall
be discussed fully below.  The three families of quarks and
leptons are placed in the following representations of $G$:
\begin{equation}
\begin{array}{c}
Q_L \sim \left( 3,2,\frac{1}{3} \right) \left( \3 \right) \\
\\
u_R \oplus u'_R \oplus u''_R \sim \left( 3,1,\frac{4}{3} \right)\left(\s \oplus \spr \oplus \sppr \right) \\
\\
d_R \oplus d'_R \oplus d''_R \sim \left( 3,1,-\frac{2}{3}
\right)\left(\s \oplus \spr \oplus \sppr \right)
\end{array}
\quad\
\begin{array}{c}
\ell_L \sim \left( 1,2,-1 \right) \left( \3 \right) \\
\\
\nu_R \sim \left( 1,1,0 \right)\left( \3 \right) \\
\\
e_R \oplus e'_R \oplus e''_R \sim \left( 1,1,-2 \right)\left(\s
\oplus \spr \oplus \sppr \right)
\end{array}
\end{equation}
where the $A_4$ notation is explained in the App.\ A, and the
$G_{SM}$ notation is standard. Models with similar $A_4$
assignments for the leptons and Higgs fields have been considered
with a different emphasis in Ref.\cite{a4ma}. Notice that the
right-handed neutrinos are assigned to a $\3$, whereas the
right-handed charged-fermions are each given a $\s \oplus \spr
\oplus \sppr$ structure.  The Higgs field assignments are
\begin{equation}
\Phi \sim \left( 1,2,-1 \right) \left( \3 \right),\quad \phi \sim
\left( 1,2,-1 \right) \left( \s \right),\quad \chi \sim \left(
1,1,0 \right) \left( \3 \right).
\end{equation}

The $G \otimes X$ invariant Yukawa Lagrangian is
\begin{eqnarray}
{\cal L}_{{\rm Yuk}}  = & \lambda_u & ( \overline{Q}_L \Phi
)_{\s}\, u_R + \lambda_u' ( \overline{Q}_L \Phi )_{\spr}\, u''_R
+ \lambda_u'' ( \overline{Q}_L \Phi )_{\sppr}\, u'_R + \nonumber\\
& + &  \lambda_d ( \overline{Q}_L \tilde{\Phi})_{\s}\, d_R +
\lambda'_d ( \overline{Q}_L \tilde{\Phi} )_{\spr}\, d''_R
+ \lambda''_d ( \overline{Q}_L \tilde{\Phi} )_{\sppr}\, d'_R + \nonumber\\
& + & \lambda_{\nu} ( \overline{\ell}_L \nu_R )_{\s}\, \phi + M [
\overline{\nu}_R (\nu_R)^c ]_{\s}
+ \lambda_{\chi} [ \overline{\nu}_R (\nu_R)^c ]_{\3s} \cdot \chi + \nonumber\\
& + & \lambda_e ( \overline{\ell}_L \tilde{\Phi})_{\s}\, e_R +
\lambda'_e ( \overline{\ell}_L \tilde{\Phi} )_{\spr}\, e''_R +
\lambda''_e ( \overline{\ell}_L \tilde{\Phi} )_{\sppr}\, e'_R +
h.c. \label{eq:Yuk}
\end{eqnarray}
where $\tilde{\Phi} \equiv i\tau_2 \Phi^*$.  This rather
busy-looking equation actually has a quite simple structure.  Each
charged fermion sector has three independent Yukawa terms, all
involving the $A_4$ triplet Higgs field $\Phi$ but not the
flavour-singlet $\phi$.  By construction, the neutrino sector is
different.  The neutrino Dirac term is governed by a single
coupling constant and involves $\phi$, while the right-handed
Majorana sector contains one bare Majorana mass $M$ and a single
Yukawa coupling term to the Higgs electroweak-singlet $\chi$
(which is an $A_4$ triplet).\footnote{Note that the $\3a$ product
of $\overline{\nu}_R$ and $(\nu_R)^c$ identically vanishes.}  All
told, there are only twelve (a priori complex) parameters to
describe the masses and mixings of nine Dirac and six Majorana
fermions. The restrictions will prove to be rather interesting.

The Yukawa Lagrangian of Eq.\ \ref{eq:Yuk} has the additional
symmetry $U(1)_X$, where $\ell_L$, $e_R$, $e'_R$, $e''_R$ and
$\phi$ carry $X = 1$, while all other fields have $X = 0$. This
non-flavour symmetry ensures that the $G_{SM} \otimes A_4$
invariant Yukawa term $\overline{\ell}_L \nu_R \Phi$ is absent
from the Lagrangian. Since $U(1)_X$ is anomalous, it cannot be
gauged.  The Goldstone boson arising from spontaneous $U(1)_X$
breaking through $\langle\phi\rangle \neq 0$ is phenomenologically
disallowed, so we will ultimately have to break $U(1)_X$
explicitly down to a discrete subgroup that is sufficient to
prevent the unwanted Yukawa term (see later).

Writing out the charged-fermion $f=u,d,e$ Yukawa invariants
explicitly using the rules \ref{eq:33tos}-\ref{eq:33tosppr} in the
appendix, one finds that each of the three mass matrix terms has
the form
\begin{equation}
\left( \begin{array}{ccc}
\overline{f}_{1L},\overline{f}_{2L},\overline{f}_{3L}
\end{array} \right)
\left( \begin{array}{ccc}
\lambda v_1 & \ \ \lambda' v_1 & \ \ \lambda'' v_1 \\
\lambda v_2 & \ \ \omega \lambda' v_2 & \ \ \omega^2 \lambda'' v_2 \\
\lambda v_3 & \ \ \omega^2 \lambda' v_3 & \ \ \omega \lambda'' v_3
\end{array} \right)
\left( \begin{array}{c} f_R \\ f''_R \\ f'_R
\end{array} \right) + h.c.
\end{equation}
where $\langle\Phi^0\rangle = (v_1,v_2,v_3)$ is the vacuum
expectation value (VEV) pattern for $\Phi$, the $v_i$ are taken to
be relatively real, and the $\lambda$'s have a suppressed
subscript $f$.  The numerical subscripts $1,2,3$ denote $A_4$
components, as in the appendix.

For the special VEV pattern
\begin{equation}
v_1 = v_2 = v_3 \equiv v \label{eq:C3vac}
\end{equation}
each of these mass matrices $M_f$ factorises as per
\begin{equation}
M_f = U(\omega) \left( \begin{array}{ccc} \sqrt{3}\lambda_f v & 0
& 0 \\ 0 & \sqrt{3} \lambda'_f v & 0 \\ 0 & 0 & \sqrt{3}
\lambda''_f v
\end{array} \right),
\label{eq:Mf}
\end{equation}
so that the left-diagonalisation matrices $V_L^{u,d,e}$ for,
respectively, the upquark, down-quark and charged-lepton sectors
are identical and equal to the unitary ``trimaximal mixing
matrix''
\begin{equation}
U(\omega) = \frac{1}{\sqrt{3}} \left( \begin{array}{ccc} 1 & 1 & 1
\\ 1 & \omega & \omega^2 \\ 1 & \omega^2 & \omega
\end{array} \right).
\end{equation}
Notice that all nine mass eigenvalues are a priori arbitrary,
despite the totally prescribed diagonalisation matrices.  This is
an example of ``form diagonalisability'', a term coined in
Ref.\cite{lowvolkas} to describe exactly this situation.
The process here is a complete contrast to the popular strategy of
relating mixing angles to mass ratios.

One immediately finds that, at this order, the chosen $A_4$
structure of the field content and the $\langle\Phi\rangle$ vacuum
forces the CKM matrix to be the identity:
\begin{equation}
V_{CKM} = V_L^{d\dagger} V_L^u = U(\omega)^{\dagger} U(\omega) =
1.
\end{equation}
The vacuum is a very special one, as it induces the breakdown
\begin{equation}
A_4 \to Z_3 \cong C_3 = \{1,c,a\},
\end{equation}
where $\cong$ denotes ``isomorphism''. The flavour group is not
broken completely at this stage, but only to the three-fold
subgroup that cyclically permutes the three $A_4$ triplet basis
states without changing their signs [see Eq.\ \ref{eq:ca}]. The
$\spr$ and $\sppr$ spaces transform under this subgroup exactly as
they do under the full group $A_4$.  As we show below, the $C_3$
remnant, if forever unbroken, is powerful enough to ensure that
the CKM matrix remains trivial to all orders.

Now to one of our main points: It is quite possible that the
reason why the observed CKM matrix is nearly the identity is the
hierarchical breaking
\begin{equation}
A_4 \to C_3 \to {\rm nothing},
\end{equation}
with the small mixing angles generated by higher-order effects
after the relatively weak subsequent breaking of the residual
$C_3$.  Before taking this line of thought further, we need to
examine the neutrino sector.

The neutrino Dirac mass matrix is different from that of the
charged-leptons, being derived from the Yukawa term
$\overline{\ell}_L \nu_R \phi$, where the fermion bilinear sees
the two $A_4$ triplets coupling to the singlet.  From Eq.\
\ref{eq:33tos}, one simply gets that the Dirac mass matrix is
proportional to the $3 \times 3$ identity matrix,
\begin{equation}
M_\nu^D = \lambda_\nu\, v_\phi\, 1 \equiv m_\nu^D\, 1,
\end{equation}
where $\langle\phi^0\rangle = v_{\phi}$. The right-handed neutrino
bare Majorana mass term is similarly trivial, being $M$ times the
identity.  The required non-trivial structure is supplied by the
Yukawa coupling to $\chi$, which expanded out is
\begin{equation}
\lambda_\chi \ \left( \begin{array}{ccc} \overline{\nu}_{1R},
\overline{\nu}_{2R}, \overline{\nu}_{3R} \end{array} \right)
\left( \begin{array}{ccc} 0 & \chi_3 & \chi_2 \\ \chi_3 & 0 &
\chi_1 \\ \chi_2 & \chi_1 & 0
\end{array} \right)
\left( \begin{array}{c} (\nu_{1R})^c \\ (\nu_{2R})^c \\
(\nu_{3R})^c \end{array} \right).
\end{equation}

We now make our second key assumption about $A_4$ breaking:  we
want
\begin{equation}
\langle\chi_1\rangle = \langle\chi_3\rangle = 0,\quad
\langle\chi_2\rangle \equiv v_{\chi} \neq 0, \label{eq:Z2vac}
\end{equation}
so that the full $6 \times 6$ neutrino mass matrix is
\begin{equation}
\left( \begin{array}{cccccc}
0 & 0 & 0 & m_{\nu}^D & 0 & 0 \\
0 & 0 & 0 & 0 & m_{\nu}^D & 0 \\
0 & 0 & 0 & 0 & 0 & m_{\nu}^D \\
m_{\nu}^D & 0 & 0 & M & 0 & M_\chi \\
0 & m_{\nu}^D & 0 & 0 & M & 0 \\
0 & 0 & m_{\nu}^D & M_\chi & 0 & M
\end{array} \right),
\end{equation}
where $M_\chi \equiv \lambda_\chi v_\chi$.  Note that $M$ and
$M_\chi$ are in general complex numbers with a relative phase
difference.  In the see-saw limit $|M|, |M_\chi| \gg m_{\nu}^D$,
the effective $3 \times 3$ mass matrix $M_L$ for the light
neutrino sector is simply
\begin{equation}
M_L = - M_\nu^D M_R^{-1} (M_\nu^D)^T = - \frac{(m_\nu^D)^2}{M}
\left( \begin{array}{ccc}
\frac{M^2}{M^2-M^2_\chi}  & 0 & - \frac{M M_\chi}{M^2-M^2_{\chi}} \\
0 & 1 & 0 \\
- \frac{M M_\chi}{M^2-M^2_{\chi}} & 0 & \frac{M^2}{M^2-M^2_\chi}
\end{array} \right),
\label{eq:MLbare}
\end{equation}
whose diagonalisation matrix is simply
\begin{equation}
V_L^{\nu} = \frac{1}{\sqrt{2}} \left( \begin{array}{ccc} 1 & 0 &
-1 \\ 0 & \sqrt{2} & 0 \\ 1 & 0 & 1
\end{array} \right).
\end{equation}
The MNSP matrix, at this order, is then
\begin{equation}
V_{MNSP} = V_L^{e\dagger} V_L^{\nu} = U(\omega)^{\dagger}
V_L^{\nu}
 = \left( \begin{array}{ccc}
\frac{2}{\sqrt{6}} & \frac{1}{\sqrt{3}} & 0 \\
-\frac{\omega}{\sqrt{6}} & \frac{\omega}{\sqrt{3}} & -\frac{e^{i\pi/6}}{\sqrt{2}} \\
-\frac{\omega^2}{\sqrt{6}} & \frac{\omega^2}{\sqrt{3}} &
\frac{e^{-i\pi/6}}{\sqrt{2}}
\end{array} \right)
\end{equation}
which, up to phases, is tribimaximal and hence fits the neutrino
oscillation data well.  These neutrino results are the same as in
the recently explored scenario of Ref.\ \cite{a4he}; we refer
the reader to this paper for a more extended phenomenological
discussion.

In the neutrino sector, the flavour breaking pattern driven by
$\langle\chi\rangle$ is
\begin{equation}
A_4 \to Z_2 = \{1,r_2\}.
\end{equation}
This $Z_2$ subgroup does not commute with the $C_3$ subgroup of
the charged-fermion sector. The neutrino and charged-fermion
sectors form ``parallel worlds'' of flavour symmetry breaking.
These parallel symmetry breaking worlds cannot be sequestered from
each other completely, of course. For the theory as a whole, $A_4$
is completely broken.

After flavour symmetry breaking, higher-order and radiative
effects will in general create terms that violate $A_4$.  We will
divide these higher-order effects into two classes:  those that
involve effects {\em within each sector}, and those that involve
{\em interactions between the two sectors}.  The former are
precisely the effects that preserve $C_3$ and $Z_2$, respectively,
for the charged-lepton and neutrino sectors.  The latter violate
$A_4$ completely.  We wish to see how these different classes
correct the CKM amd MNSP matrices from the trivial and
tribimaximal forms, respectively.  We shall work in as
dynamics-independent a way as possible.

\section{Corrections within each sector after flavour symmetry breaking.}

Let us start with the charged-fermion sector.  At the bare
Lagrangian level, the only Yukawa terms allowed are those
invariant under $A_4$.  After $A_4$ spontaneously breaks to $C_3$
in this sector, higher-order effects will generate Yukawa terms
that violate $A_4$ but respect $C_3$. We now write down all those
terms.

Under $C_3 = \{1,c,a\}$, the triplets $Q_L$, $\ell_L$ and $\Phi$
transform as per
\begin{equation}
c: (1,2,3) \to (3,1,2)\quad {\rm and}\quad a: (1,2,3) \to (2,3,1),
\end{equation}
where $1,2,3$ denote the triplet entries, as before.  The $A_4$
singlets $f_R$ become $C_3$ singlets ($f=u,d,e$ as before), while
the non-trivial one-dimensional $A_4$-plets $f'_R$ and $f''_R$
transform thus:
\begin{equation}
f'_R \left\{ \begin{array}{c} \stackrel{c}{\to} \omega\ f'_R \\
\stackrel{a}{\to} \omega^2 f'_R
\end{array} \right. \quad {\rm and}\quad
f''_R \left\{ \begin{array}{c} \stackrel{c}{\to} \omega^2 f''_R \\
\stackrel{a}{\to} \omega\ f''_R
\end{array} \right.
\end{equation}
The previously allowed $A_4$ invariant $(\overline{f}_{1L}\Phi_1^0
+ \overline{f}_{2L}\Phi_2^0 +\overline{f}_{3L}\Phi_3^0)\ f_R$
(where the $\Phi_i^0$ generically denote the charge-neutral fields
within $\Phi$ and $\tilde{\Phi}$) is now supplemented with the
following terms that violate $A_4$ but respect $C_3$:
\begin{eqnarray}
& (\overline{f}_{1L}\Phi_2^0 + \overline{f}_{2L}\Phi_3^0 +
\overline{f}_{3L}\Phi_1^0)\ f_R &
\nonumber\\
& (\overline{f}_{1L}\Phi_3^0 + \overline{f}_{2L}\Phi_1^0 +
\overline{f}_{3L}\Phi_2^0)\ f_R. &
\end{eqnarray}
Similarly, the $A_4$ invariants $(\overline{f}_{1L}\Phi_1^0 +
\omega \overline{f}_{2L}\Phi_2^0 + \omega^2
\overline{f}_{3L}\Phi_3^0)\ f''_R$ and $(\overline{f}_{1L}\Phi_1^0
+ \omega^2 \overline{f}_{2L}\Phi_2^0 + \omega
\overline{f}_{3L}\Phi_3^0)\ f'_R$ are joined by
\begin{eqnarray}
& (\overline{f}_{1L}\Phi_2^0 + \omega \overline{f}_{2L}\Phi_3^0 +
\omega^2 \overline{f}_{3L}\Phi_1^0)\ f''_R &
\nonumber\\
& (\overline{f}_{1L}\Phi_3^0 + \omega \overline{f}_{2L}\Phi_1^0 +
\omega^2 \overline{f}_{3L}\Phi_2^0)\ f''_R &
\nonumber\\
& (\overline{f}_{1L}\Phi_2^0 + \omega^2 \overline{f}_{2L}\Phi_3^0
+ \omega \overline{f}_{3L}\Phi_1^0)\ f'_R &
\nonumber\\
& (\overline{f}_{1L}\Phi_3^0 + \omega^2 \overline{f}_{2L}\Phi_1^0
+ \omega \overline{f}_{3L}\Phi_2^0)\ f'_R. &
\end{eqnarray}
Each of the new terms comes, generically, with a different
coupling constant.  It is interesting, though, that despite all
these new Yukawa terms, the mass matrices retain the form of Eq.\
\ref{eq:Mf} once the $C_3$-preserving VEV pattern
$\langle\Phi_1^0\rangle = \langle\Phi_2^0\rangle =
\langle\Phi_3^0\rangle \equiv v$ is used.  This means that the
left-diagonalisation matrices for the $u$, $d$ and $e$ sectors
remain trimaximal, and hence the CKM matrix remains trivial.  We
have thus demonstrated that it is the $C_3$ subgroup of $A_4$ that
is responsible for preventing quark mixing.  The origin of CKM
mixing must then arise from $C_3$ breaking, which in the spirit of
economy one may wish to extract from the neutrino sector (though
this is not mandatory -- one may also extend the theory).

We now turn to the neutrino sector. It is easy to see that the
minus sign associated with the unbroken $Z_2$ transformations
keeps the $(1,2)$, $(2,1)$, $(2,3)$ and $(3,2)$ entries of both
the neutrino Dirac and right-handed Majorana mass matrices zero.
However, these two matrices need not be proportional to the
identity any longer, as that feature was driven by the now broken
$A_4$.  We now have, in general, that
\begin{equation}
M_{\nu}^D = \lambda_{\nu}\, \langle\phi\rangle\, \left. \left(
\begin{array}{ccc} 1 & 0 & 0 \\ 0 & 1 & 0 \\ 0 & 0 & 1
\end{array} \right) \right|_{\rm bare} +
\left. \left( \begin{array}{ccc} \epsilon_{11} & 0 & \epsilon_{13}
\\ 0 & \epsilon_{22} & 0 \\ \epsilon_{31} & 0 & \epsilon_{33}
\end{array} \right) \right|_{\rm h.o.}
\end{equation}
and the bare right-handed Majorana mass matrix is
\begin{equation}
M \, \left. \left( \begin{array}{ccc} 1 & 0 & 0 \\ 0 & 1 & 0 \\ 0
& 0 & 1
\end{array} \right) \right|_{\rm bare} +
\left. \left( \begin{array}{ccc} \epsilon'_{11} & 0 &
\epsilon'_{13} \\ 0 & \epsilon'_{22} & 0 \\ \epsilon'_{31} & 0 &
\epsilon'_{33}
\end{array} \right) \right|_{\rm h.o.}
\end{equation}
where subscript ``h.o.'' stands for ``higher order'', with the
$\epsilon$ and $\epsilon'$ entries being small.  The $\nu_R-\chi$
coupling terms now also contain the independent $Z_2$ invariants
\begin{eqnarray}
& \overline{\nu}_{2R} (\nu_{3R})^c \chi_{1,3},\ \
\overline{\nu}_{3R} (\nu_{1R})^c \chi_2,\ \
\overline{\nu}_{1R} (\nu_{2R})^c \chi_{3,1}, &\nonumber\\
& \overline{\nu}_{1R} (\nu_{1R})^c \chi_2,\ \ \overline{\nu}_{2R}
(\nu_{2R})^c \chi_2,\ \ \overline{\nu}_{3R} (\nu_{3R})^c \chi_2. &
\end{eqnarray}
Inputting the $Z_2$-preserving vacuum of Eq.\ \ref{eq:Z2vac}, we
see that the new terms involve corrections to the $(i,i)$ and
$(1,3)=(3,1)$ entries in the right-handed Majorana mass matrix. In
total then, we have that the right-handed Majorana mass terms are
\begin{equation}
\left. \left( \begin{array}{ccc} M & 0 & M_\chi \\ 0 & M & 0 \\
M_\chi & 0 & M
\end{array} \right) \right|_{\rm bare} +
\left. \left( \begin{array}{ccc} \epsilon''_{11} & 0 &
\epsilon''_{13} \\ 0 & \epsilon''_{22} & 0 \\ \epsilon''_{31} & 0
& \epsilon''_{33}
\end{array} \right) \right|_{\rm h.o.}
\end{equation}
It is obvious then that the effective $M_L$ is additively
corrected from Eq.\ \ref{eq:MLbare} by
\begin{equation}
M_L \to M_L + \left. \left( \begin{array}{ccc} \delta_{11} & 0 &
\delta_{13} \\ 0 & \delta_{22} & 0 \\ \delta_{13} & 0 &
\delta_{33}
\end{array} \right) \right|_{\rm h.o.},
\end{equation}
where, in general, the $\delta_{ij}$ are complex. The neutrino
left-diagonalisation matrix is now corrected to
\begin{equation}
V_L^{\nu} = \left( \begin{array}{ccc} 1 & 0 & 0 \\ 0 & 1 & 0 \\ 0
& 0 & e^{i\beta}
\end{array} \right)
\left( \begin{array}{ccc} \cos\theta & \ 0 & \ -\sin\theta \\ 0 &
\ 1 & \ 0 \\ \sin\theta & \ 0 & \ \cos\theta
\end{array} \right)
\left( \begin{array}{ccc} e^{i\alpha_1} & 0 & 0 \\ 0 &
e^{i\alpha_2} & 0 \\ 0 & 0 & e^{i\alpha_3}
\end{array} \right),
\end{equation}
where
\begin{equation}
\theta = \frac{\pi}{4} + \delta
\end{equation}
and $\delta \ll 1$.  The phases $\alpha_i$ can be absorbed into
the neutrino mass eigenstate fields, but the phase $\beta$, given
by
\begin{equation}
\beta = {\rm Arg}(M + \delta_{33}) - {\rm Arg}(M + \delta_{11}),
\end{equation}
is important because it will contribute to $CP$ violation in
neutrino oscillations.

The MNSP matrix becomes
\begin{equation}
V_{MNSP} = U(\omega)^{\dagger} V_L^{\nu} = \frac{1}{\sqrt{3}}
\left( \begin{array}{ccc}
c + s e^{i\beta} & 1 & c e^{i\beta} - s \\
c + \omega s e^{i\beta} & \omega^2 & \omega c e^{i\beta} - s \\
c + \omega^2 s e^{i\beta} & \omega & \omega^2 c e^{i \beta} - s
\end{array} \right),
\end{equation}
where $c \equiv \cos\theta$ and $s \equiv \sin\theta$.
The middle column is uncorrected at this level, a nonzero $U_{e3}$
element is generated, and there are other small deviations from
exact tribimaximal mixing.

$CP$ violation in neutrino oscillations is generated at this
level. The Jarlskog invariant is
\begin{equation}
{\rm Im}[V_{11}\, V_{12}^*\, V_{21}^*\, V_{22}] = \frac{1}{9}
(\cos 2\theta - \sin 2\theta \sin\beta)
\sin\left(\frac{2\pi}{3}\right)
\end{equation}
where the $V_{ij}$ denote the entries of $V_{MNSP}$.

\section{Interactions between the sectors after flavour symmetry breaking}

So far, we have neglected interactions between the charged-fermion
(plus $\Phi$) sector and the neutrino (plus $\chi$ and $\phi$)
sector, the two parallel worlds of flavour symmetry breaking.  The
neutrino-world $A_4$ breaking is sufficient to generate realistic
neutrino mass and mixing phenomenology while at the same time
explaining why the dominant mixing pattern is tribimaximal.  The
charged-fermion world, however, has an interesting residual $C_3$
symmetry that prevents the generation of quark mixing.  This is
pleasing at lowest order because of the known fact that CKM angles
are small; clearly, however, the $C_3$ subgroup must be slightly
broken to achieve a fully realistic quark sector.

As remarked earlier, the full theory does in fact have broken
$C_3$, as the neutrino sector does not respect it.  This suggests
that one should look to $C_3$ breaking mediated to the quark
sector from the neutrino/$\chi$ sector as the natural source for
quark mixing, which we may term ``cross-talk''.

The details of the cross-talk depend on the specific dynamics.
Since we want to use symmetry on its own as much as possible, we adopt an
effective operator approach.\footnote{Note that the analysis of
the preceding section could also have been phrased in the language
of effective operators.  For instance, the $\nu_R - \chi$ terms
that violate $A_4$ but preserve $Z_2$ can originate from
higher-dimensional operators such as $\overline{\nu}_R (\nu_R)^c
\chi^{n+1}$ where $n$ of the $\chi$'s are replaced by their VEVs.}
This allows us to identify what the symmetry breakdown structure
in principle allows by the way of dynamical outcomes.  If realistic
quark mixing were to be allowed, then that would motivate the
construction of explicit models.

The effective operators relevant for quark mixing are at
dimension-five.  Schematically, they are of the forms
\begin{eqnarray}
& \overline{Q}_L \, u_R \, \Phi \, \chi,\quad \overline{Q}_L \,
u'_R \, \Phi \, \chi,\quad \overline{Q}_L \, u''_R \, \Phi \, \chi
&
\nonumber\\
& \overline{Q}_L \, d_R \, \tilde{\Phi} \, \chi,\quad
\overline{Q}_L \, d'_R \, \tilde{\Phi} \, \chi,\quad
\overline{Q}_L \, d''_R \, \tilde{\Phi} \, \chi, &
\label{eq:quarkeffops}
\end{eqnarray}
plus similar operators with $\Phi \to \phi$.  The VEV of $\chi$
communicates $C_3$ breaking to the quarks through these operators.
The operators involving $\phi$ are eliminated by the $U(1)_X$
symmetry (or the relevant discrete subgroup thereof), so we need
only consider the set in Eq.\ \ref{eq:quarkeffops}.  Each operator
is suppressed either by a high mass scale $M_{\it inter}$
that characterises the dynamical interactions between the two
sectors (or a set of such scales), or by small coupling constants
controlling those interactions.

Looked at more carefully, we see that each of the above yields two
independent $A_4$ invariants.  For example, $\overline{Q}_L \, u_R
\, \Phi \, \chi$ schematically denotes the independent terms
\begin{equation}
[\, (\, \overline{Q}_L\, \Phi\, )_{\3s}\, \chi\, ]_{\s}\, u_R\quad
{\rm and}\quad [\, (\, \overline{Q}_L\, \Phi\, )_{\3a}\, \chi\,
]_{\s}\, u_R. \label{eq:35}
\end{equation}
Expanding out these terms, and inserting the VEVs of Eqs.\
\ref{eq:C3vac} and \ref{eq:Z2vac}, it is easy to see that the
dimension-five operators yield corrections to the quark mass
matrices of the forms
\begin{equation}
\Delta M_{u,d} = \left( \begin{array}{ccc}
x_1^{u,d} & \ \ x_2^{u,d} & \ \ x_3^{u,d} \\
0 & \ \ 0 & \ \ 0 \\
y_1^{u,d} & \ \ y_2^{u,d} & \ \ y_3^{u,d}
\end{array} \right)
\end{equation}
where the entries are in general complex.  The corrected mass
matrices are then easily seen to be
\begin{eqnarray}
M + \Delta M & = & U(\omega) \, \sqrt{3}\, \left(
\begin{array}{ccc}
\lambda v + (x_1 + y_1)/3 & \ \ (x_2 + y_2)/3 & \ \ (x_3 + y_3)/3 \\
(x_1 + \omega y_1)/3 & \ \ \lambda' v + (x_2 + \omega y_2)/3 & \ \ (x_3 + \omega y_3)/3 \\
(x_1 + \omega^2 y_1)/3 & \ \ (x_2 + \omega^2 y_2)/3 & \ \
\lambda'' v + (x_3 + \omega^2 y_3)/3
\end{array} \right) \nonumber\\
& \equiv & U(\omega) V_L \left( \begin{array}{ccc} m_{u,d} & 0 & 0
\\ 0 & m_{c,s} & 0 \\ 0 & 0 & m_{t,b}
\end{array} \right) V_R^{\dagger}
\end{eqnarray}
where some of the $u,d$ labels have, for clarity, been suppressed.
The left-diagonalisation matrices are now $U(\omega) V_L^{u,d}$
instead of just $U(\omega)$, where $V_L^{u,d}$ are nearly diagonal
if the $x$'s and $y$'s are smaller in magnitude than the $\lambda
v$'s.  This means that the CKM matrix is now given by
\begin{equation}
V_{CKM} = [\, U(\omega)\, V_L^d\, ] ^{\dagger}\, [\, U(\omega)\,
V_L^u\, ] = V_L^{d\dagger}\, V_L^u \neq 1.
\end{equation}
In general, there is enough freedom in $V_L^{u,d}$ to fit the
observed CKM matrix, while also, of course, explaining why it is
nearly the identity.

So, we conclude that the overall structure allows the
$\chi$-sector to seed the $C_3$ breaking required to generate
quark mixing.  To understand more detailed features of the CKM
matrix, such as the relative magnitudes of the $(1,2)$, $(2,3)$
and $(1,3)$ entries, it appears one would need an explicit
fundamental theory with the correct relative sizes for the $x$'s
and $y$'s.

Generically, one also expects the charged-fermion-$\Phi$ sector to
feed through into the neutrino sector and hence to provide
additional deviation from tribimaximal mixing.  The extent to
which this would happen is model-dependent.  It would be pleasing
to construct a model where these effects were actually very small,
in order to preserve the appealing neutrino mixing pattern
described in the previous section.

\section{Ideas for an underlying dynamics}

We have described an approach to understanding flavour mixing angles
driven as much as possible purely by spontaneous $A_4$ symmetry
breaking and its generic consequences.  We made limited
use of the language of Higgs fields and their expectation values
for concreteness and convenience, but otherwise tried to be
dynamically non-committed.

Eventually, though, one wants a dynamical completion for our
scenario.  The possibilities for this may only be limited by the
creative powers of the model builder.  However, there is an important
challenge: it is not trivial to ensure that the different vacuum
alignments of $\langle\Phi\rangle$ and $\langle\chi\rangle$,
as per Eqs.\ \ref{eq:C3vac} and \ref{eq:Z2vac}, are preserved, or at
least approximately preserved.
Since these VEV patterns are absolutely fundamental to our scheme,
the alignment problem is obviously an important one.
In this section, we shall briefly discuss this challenge and suggest
possible solutions.

First, let us study the most straightforward dynamical completion: a
standard renormalisable Higgs potential.  The full quartic $G_{SM} \otimes
A_4$-invariant Higgs potential in $\Phi$, $\phi$ and $\chi$
is displayed in App.\ B. As expected, it has a large number of
terms (more than two dozen) and so may not be compelling as a
serious model for nature. Nevertheless, it is amenable to
analysis, and it highlights the vacuum alignment challenge.

In the spirit of the ``parallel symmetry-breaking worlds'' paradigm,
it is useful to think of the Higgs potential as the sum of several pieces,
\begin{equation}
V = V(\Phi) + V(\chi) + V(\phi) + V(\Phi,\chi) + V(\Phi,\phi) +
V(\phi,\chi) + V(\Phi,\chi,\phi),
\end{equation}
where the first three terms are the self-interactions of the three
Higgs multiplets, while the remaining terms describe interactions
between them in an obvious notation.

It is easy to check that Eq.\ \ref{eq:C3vac} can be a global minimum of
$V(\Phi)$, and that Eq.\ \ref{eq:Z2vac} can be a global minimum of
$V(\chi)$ (and, of course, $V(\phi)$ is manifestly well-behaved
since $\phi$ is a flavour singlet). In all cases, the required Higgs potential
parameter space is large.

This is no longer the situation once interactions between $\Phi$ and $\chi$ are switched
on via $V(\Phi,\chi)$.  The problem is that the extremum conditions furnish
a larger number of independent equations than there are unknown VEVs ($v$, $v_\chi$ and $v_\phi$).
This means that unnatural fine-tuning conditions have to be enforced on the Higgs
potential parameters.  The troublesome interaction terms all reside in $V(\Phi,\chi)$,
so the most straightforward approach to solving the problem is to find models
in which these interactions naturally vanish.\footnote{It is worth noting that nonzero
values for $\lambda_1^{\Phi\chi}$ and $\lambda_4^{\Phi\chi}$ only are consistent
with the extremisation conditions, so having a completely vanishing $V(\Phi,\chi)$
may not be strictly necessary.}

Before suggesting natural ways to make $V(\Phi,\chi)=0$,
we very briefly digress to highlight the importance of the $\Phi-\phi$ interaction term
\begin{equation}
\lambda_3^{\Phi\phi}(\Phi^{\dagger} \phi) \cdot (\Phi^{\dagger} \phi) + h.c.
\end{equation}
If it is nonzero, then the additional
global symmetry $U(1)_X$ is explicitly broken to a $Z_2$ subgroup
under which $\phi \to -\phi$, and $\ell_L$ and all three
right-handed charged-lepton fields also change sign.  This
explicit breaking removes the phenomenologically dangerous
potential Goldstone boson, while leaving a discrete subgroup to
ensure the absence of the $\overline{\ell}_L \nu_R \Phi$ Yukawa
term.

One way to solve the vacuum alignment problem is to introduce additional symmetries
to enforce $V(\Phi,\chi) = 0$.  The difficulty here is that the most obvious candidate
transformations cannot enforce this condition.  For the case of a real $\chi$ field, the
only additional internal transformation allowed is simply $\chi \to -\chi$.  But terms of
the form $\Phi^{\dagger}\Phi\chi^2$ are always invariant under this transformation.  Similarly,
$\Phi \to e^{i\theta}\Phi$ transformations are always invariances.  If the model is extended
by making $\chi$ complex, the natural transformation $\chi \to e^{i\alpha} \chi$ cannot forbid
$\Phi^{\dagger}\Phi \chi^{\dagger}\chi$ terms.

We have thought of two generic symmetry principles that have potential application to
this problem.  They both involve spacetime transformations, as purely internal ones do not appear to
have sufficient power, as explained above.

The first possibility is to consider the limit
where $\chi$ completely decouples from the rest of the fields, because (in the absence
of gravity) the theory is then invariant under independent Lorentz transformations for $\chi$,
on the one hand, and the rest of the fields, on the other.\footnote{This fact may be unfamiliar,
though it has been shown to be of relevance for the invisible axion model \cite{dv}.  It is clear
that if all interaction terms between $\chi$ and everything else are switched off, then they
cannot be generated radiatively.  Therefore, in the absence of gravity, there should be an
increase in the symmetry of the theory as the $\chi$-decoupling limit is taken.  That symmetry
increase is for Lorentz transformations of $\chi$ to be independent of those for all the other fields.}
The decoupling of $\chi$ is achieved in the limit
\begin{equation}
\lambda_\chi \to 0,\qquad \lambda^{\phi\chi} \to 0.
\end{equation}
We do not, however, wish $\lambda_\chi$ to be precisely zero, otherwise the neutrinos would be exactly
degenerate.  If $\lambda_\chi$ is small but nonzero, it might be possible to generate an acceptable
neutrino mass spectrum while radiatively inducing sufficiently small $V(\Phi,\chi)$ terms, where
``sufficiently small'' means that those terms alter the required VEV pattern by only a small amount, hence
preserving the benefits of that symmetry breaking pattern.  A detailed analysis of this possibility
is beyond the scope of this paper.

The second spacetime symmetry principle is supersymmetry, acting in concert with internal
symmetries.  Quartic $V(\Phi,\chi)$ terms can only arise
from F-terms.  Since the superpotential is at most cubic, this means the generic interaction
term is of the form $\Phi_u \Phi_d \chi$, where $\Phi_{u,d}$ are the two Higgs chiral superfields
required by supersymmetry, and $\chi$ now represents the chiral superfield containing the scalar
component of the same name.  Internal transformations {\it can} forbid that cubic
superpotential term, thus also forbidding quartic $V(\Phi,\chi)$ terms.  (For definiteness, think
of $\chi \to -\chi$, though the actual transformations in a realistic supersymmetric extension
are almost certainly going to be more involved.)  The cubic terms in $V(\Phi,\chi)$ are soft supersymmetry
breaking terms such as $\Phi_{u,d}^{\dagger} \Phi_{u,d} \chi$ and $\Phi_u \Phi_d \chi$, which can
also be forbidden by suitable internal (probably discrete) symmetries such as $\chi \to -\chi$.
An attempt to construct such a supersymmetric dynamical completion is discussed in App.\ C.
It serves as an existence proof that a dynamical completion is possible.  We have not as yet
tried to optimise the model building process, a topic we hope to return to in a future
paper.

Another approach worth pursuing, very different from the above, is to take the ``parallel
worlds of symmetry breaking'' language literally, by sequestering \cite{rs} $\Phi$ and $\chi$
on different branes in an extra-dimensional setting (see Ref.\ \cite{a4alterali}
for an example of a brane-world approach to the neutrino problem).  The
physical separation of $\Phi$ and $\chi$ is another generic way to forbid the problematic
interaction terms.

The dynamical completion issue should, however, not
distract us from the main point of
this paper: the $A_4$ flavour symmetry breaking structure
outlined in the preceeding sections.

\section{Conclusion}

We have proposed an $A_4$ flavour structure that fits very well
with the observed patterns of quark and neutrino mixing, while
leaving mass eigenvalues arbitrary. The $A_4$ field content and
Higgs VEV patterns were selected to produce, at lowest order,
neutrino tribimaximal mixing and zero quark mixing.  The required
structure splits naturally into two sectors: the
neutrino/$\chi$/$\phi$ domain and the charged-fermion/$\Phi$
domain.  Different spontaneous flavour breaking patterns occur in
these parallel worlds of symmetry breaking.  The charged-fermion
sector has $A_4 \to C_3$, while the neutrinos see $A_4 \to Z_2$.
Radiative or higher-order effects {\em within} each sector that
violate $A_4$ but preserve the respective subgroup were examined.
The $C_3$ symmetry prevents the generation of quark mixing, while
a small and interesting deviation from tribimaximal mixing is
allowed by the $Z_2$.  The latter includes a small but nonzero
$U_{e3}$ and has $CP$ violation in neutrino oscillations.  For
quark mixing to be induced, $C_3$ has to be broken.  We explored
the natural possibility that the neutrino sector communicates its
$C_3$ breaking to the quarks, and showed through an effective
operator analysis that generically this does induce a realistic
CKM matrix.  Finally, we discussed the dynamical completion challenge
and supplied an existence proof that it can be met.  Finding the most elegant
possible underlying dynamics remains a goal for the future.

We believe that the proposed flavour symmetry structure is a
promising base from which to explore the fundamental origin of
quark and lepton mixing angles.

\acknowledgments RRV would like to thank the NCTS/TPE in the
Department of Physics at National Taiwan University for
hospitality while part of this work was carried out, and A.
Coulthurst, A. Demaria, J. Doukas, K. McDonald and A. Zee for
helpful conversations. RRV was also partially supported by the
Australian Research Council. XGH and YYK were partially funded by
the ROC National Science Council. YYK was also partially supported
by CHEP/Kyungpook National University in Korea.

\newpage

\appendix

\section{Basic $A_4$ properties.}

The alternating group of order four, denoted $A_4$, is defined as
the set of all twelve even permutations of four objects.  It has a
real three-dimensional irreducible representation $\3$, and three
inequivalent one-dimensional representations $\s$, $\spr$ and
$\sppr$.  The representation $\s$ is trivial, while $\spr$ and
$\sppr$ are non-trivial and complex conjugates of each other.

The twelve representation matrices for $\3$ are conveniently taken
to be the $3 \times 3$ identity matrix $1$, the reflection
matrices $r_1 \equiv {\rm diag}(1,-1,-1)$, $r_2 \equiv {\rm
diag}(-1,1,-1)$ and $r_3 \equiv {\rm diag}(-1,-1,1)$, the cyclic
and anticyclic matrices
\begin{equation}
c = a^{-1} \equiv \left( \begin{array}{ccc} 0 & 0 & 1 \\ 1 & 0 & 0
\\ 0 & 1 & 0 \end{array} \right)\quad {\rm and}\quad a = c^{-1}
\equiv \left( \begin{array}{ccc} 0 & 1 & 0 \\ 0 & 0 & 1 \\ 1 & 0 &
0 \end{array} \right), \label{eq:ca}
\end{equation}
respectively, as well as $r_i c r_i$ and $r_i a r_i$.  Under the
group element corresponding to $c (a)$, $\spr \to \omega
(\omega^2) \spr$ and $\sppr \to \omega^2 (\omega) \sppr$, where
$\omega = e^{i2\pi/3}$ is a complex cube-root of unity, with both
being unchanged under the $r_i$.

The basic non-trivial tensor products are
\begin{equation}
\3 \otimes \3 = \3_s \oplus \3_a \oplus \s \oplus \spr \oplus
\sppr,\quad {\rm and}\quad \spr \otimes \spr = \sppr,
\label{eq:A4tensors}
\end{equation}
where $s (a)$ denotes symmetric (antisymmetric) product.  Let
$(x_1,x_2,x_3)$ and $(y_1,y_2,y_3)$ denote the basis vectors for
two $\3$'s. Then
\begin{eqnarray}
(\3 \otimes \3)_{\3s}  & = & ( x_2 y_3 + x_3 y_2\, ,\, x_3 y_1 +
x_1 y_3\, ,\, x_1 y_2 + x_2 y_1 ),
\label{eq:33to3s}\\
(\3 \otimes \3)_{\3a}  & = & ( x_2 y_3 - x_3 y_2\, ,\, x_3 y_1 -
x_1 y_3\, ,\, x_1 y_2 - x_2 y_1 ),
\label{eq:33to3a}\\
(\3 \otimes \3)_{\s} & = & x_1 y_1 + x_2 y_2 + x_3 y_3, \label{eq:33tos}\\
(\3 \otimes \3)_{\spr} & = & x_1 y_1 + \omega\, x_2 y_2 + \omega^2\, x_3 y_3, \label{eq:33tospr}\\
(\3 \otimes \3)_{\sppr} & = & x_1 y_1 + \omega^2\, x_2 y_2 +
\omega\, x_3 y_3, \label{eq:33tosppr}
\end{eqnarray}
in an obvious notation.

\section{Higgs Potential}

The $G_{SM} \otimes A_4$-invariant,
renormalisable Higgs potential terms consistent with the discrete $Z_2$ subgroup of
$U(1)_X$ are given by

\begin{eqnarray}
V(\Phi) &=& \mu^2_\Phi (\Phi^\dagger \Phi)_{\s} +\lambda^\Phi_1
(\Phi^\dagger \Phi)_{\s}(\Phi^\dagger \Phi)_{\s} + \lambda^\Phi_2
(\Phi^\dagger \Phi)_{\s'}(\Phi^\dagger \Phi)_{\s''}\nonumber\\
&+&\lambda^\Phi_3 (\Phi^\dagger \Phi)_{\3s}(\Phi^\dagger
\Phi)_{\3s} + \lambda^\Phi_4 (\Phi^\dagger
\Phi)_{\3a}(\Phi^\dagger \Phi)_{\3a}\nonumber\\
& +& i \lambda^\Phi_5 (\Phi^\dagger \Phi)_{\3s}(\Phi^\dagger
\Phi)_{\3a}.\\
 V(\chi) &=& \mu^2_\chi (\chi \chi)_{\s} +
\delta^\chi (\chi\chi\chi)_{\s} + \lambda^\chi_1 (\chi\chi)_{\s}
(\chi\chi)_{\s} + \lambda^\chi_2
(\chi\chi)_{\s'}(\chi\chi)_{\s''}\nonumber\\
& +& \lambda^\chi_3 (\chi\chi)_{\3}(\chi\chi)_{\3}.
\\
V(\phi) &=& \mu^2_\phi (\phi^\dagger \phi) + \lambda^\phi
(\phi^\dagger\phi)^2
\\
V(\Phi,\chi) &=& \delta^{\Phi\chi}_{s}(\Phi^\dagger \Phi)_{\3s}
\chi + i \delta^{\Phi\chi}_a (\Phi^\dagger \Phi)_{\3a}\chi+
\lambda^{\Phi\chi}_1
(\Phi^\dagger\Phi)_{\s}(\chi\chi)_{\s}\nonumber\\
&+&\lambda^{\Phi\chi}_2 (\Phi^\dagger\Phi)_{\s'}(\chi\chi)_{\s''}
+\lambda^{\Phi\chi*}_2
(\Phi^\dagger\Phi)_{\s''}(\chi\chi)_{\s'}\nonumber\\
&+&\lambda^{\Phi\chi}_3 (\Phi^\dagger\Phi)_{\3s}(\chi\chi)_{\3} +i
\lambda^{\Phi\chi}_4 (\Phi^\dagger\Phi)_{\3a}(\chi\chi)_{\3}.
\\
V(\Phi, \phi) &=& \lambda^{\Phi\phi}_1 (\Phi^\dagger\Phi)_{\s}
(\phi^\dagger \phi) + \lambda^{\Phi\phi}_2 (\Phi^\dagger
\phi)(\phi^\dagger \Phi) + \lambda^{\Phi\phi}_3 (\Phi^\dagger
\phi)(\Phi^\dagger \phi)\nonumber\\
& +& \lambda^{\Phi\phi*}_3 (\phi^\dagger \Phi)(\phi^\dagger \Phi).
\\
V(\phi, \chi) &=& \lambda^{\phi\chi} (\phi^\dagger \phi)
(\chi\chi)_{\s}.
\end{eqnarray}

There is no renormalizable term simultaneously involving $\Phi$, $\phi$ and
$\chi$ allowed by the $Z_2$ subgroup of $U(1)_X$, that is, $V(\Phi,\chi,\phi) =
0$.

The total potential is given by
\begin{equation}
V = V(\Phi) + V(\chi) + V(\phi) + V(\Phi,\chi) + V(\Phi,\phi) +
V(\phi,\chi) + V(\Phi,\chi,\phi).
\end{equation}

\section{A supersymmetric dynamical completion}

For the sake of supplying an existence theorem, we have
constructed one example of such a theory.  It is rather elaborate,
in that it requires an additional discrete $Z_{12} \otimes Z_2$
symmetry together with supersymmetry and several additional
fields. Since $\chi$ is now in a supermultiplet it becomes a
complex field, and for the usual reason the number of Higgs
doublets must be doubled.  The chiral superfield content is
\begin{eqnarray}
&&Q_L \sim(\3,1,-1),\qquad u^c_R,\ d^c_R \sim (\s \oplus \spr
\oplus \sppr,
\omega_{12}^7,-1),\nonumber\\
&&\ell_L \sim (\3,1,-1),\qquad e_R^c \sim (\s \oplus \spr \oplus
\sppr, \omega_{12}^7,-1),\qquad \nu_R \sim
(\3,\omega_{12}^4,-1),\nonumber\\
&&\Phi_{u,d} \sim (\3,\omega_{12}^{5},1),\qquad \phi_{u} \sim
(\s,\omega_{12} ^8,1),\qquad
\phi_d \sim (1, \omega_{12}^4,1)\nonumber\\
&&\chi \sim (\3,\omega_{12}^4,1),\qquad \chi' \sim (\3,\omega_{12}
^2,1),\nonumber\\
&&s \sim (\s,\omega_{12}^4,1),\qquad s' \sim
(\s,\omega_{12}^2,1),\qquad
s'' \sim (\s,\omega_{12}^8,1),
\end{eqnarray}
where $\omega_{12}^{12} = 1$. The first entry provides the $A_4$
assignment, the second specifies how that field multiplicatively
transforms under the $Z_{12}$, and the third the parity under the
$Z_2$.  The gauge assignments for the quark, lepton and
Higgs-doublet superfields are standard, with $s$, $s'$ and $s''$
being new gauge and $A_4$ singlets, while $\chi'$ is a new
gauge-singlet $A_4$ triplet.

The $Z_{12}$ charges are chosen in such a way that, first,
communication between $\chi$ and $\Phi_{u,d}$ is forbidden, so as to avoid
the troublesome terms in $V(\Phi,\chi)$, but, second, the $\chi^3$ and
$\Phi_u \Phi_d \chi'$ terms are allowed so that the desired VEV alignment can be
enforced. The $Z_2$ charge disallows terms of the type $\nu^c_R
\chi^2$ which can cause the fermion partner of $\chi$ to mix
with neutrinos and therefore destroy the pattern of the neutrino
mass matrix.

The superpotential contains the terms $Q_L \Phi_u u^c_R$,
$Q_L\Phi_d d^c_R$, $\ell_L \phi_u \nu_R^c$, $\ell_L \Phi_d e_R^c$,
$\nu^c_R \nu^c_R s$ and $\nu^c_R \nu^c_R \chi$ which supply all
the central Yukawa couplings.  There are no bare Majorana masses
for $\nu_R$, but $\nu^c_R \nu^c_R s$ generates universal Majorana
masses once the spin-0 component of $s$ acquires a VEV.

The superpotential for the Higgs multiplets is
\begin{eqnarray}
W &=& a_1 \chi^3+ a_2\chi^2s + a_3 s^3 +   a_4 \phi_u \phi_d +
a_5(\Phi_u\Phi_d)_{3s} \chi' +
a_6(\Phi_u\Phi_d)_{3a}\chi'\nonumber\\
& +& a_7\Phi_u \Phi_d s' + a_8 \chi'^2 s'' + a_9 s s''+a_{10}
s'^2s'' + a_{11}s''^3.
\end{eqnarray}
 From this structure, it is evident that all supersymmetric
$V(\Phi,\chi)$ terms are absent from the $F$-term contributions,
while the $D$-term contributions are also safe because, of course,
they cannot involve the gauge singlet $\chi$.

In the supersymmetric limit, the desired VEV structure cannot be
obtained, but supersymmetry has to be broken in any case. To this
end, we follow the usual soft supersymmetry breaking approach by adding
to the potential
all soft-breaking terms that preserve $A_4\otimes Z_{12}\otimes Z_2$.
These terms are given by
\begin{eqnarray}
V_{soft}&=&  b_1 \chi^3+ b_2\chi^2s + b_3 s^3 +   b_4 \phi_u
\phi_d + b_5(\Phi_u\Phi_d)_{3s} \chi' +
b_6(\Phi_u\Phi_d)_{3a}\chi'\nonumber\\
& +&b_7\Phi_u \Phi_d s' + b_8 \chi'^2 s'' + b_9 s s''+b_{10}
s'^2s'' +
b_{11}s''^3 + H.C.\nonumber\\
&+& c_1 \chi^\dagger \chi + c_2 s^\dagger s + c_3 s'^\dagger s'+
c_4 s''^\dagger s'' + c_5 \phi^\dagger_u \phi_u + c_6
\phi^\dagger_d \phi_d + c_7 \chi'^\dagger \chi'\nonumber\\
& + & c_8
\Phi_u^\dagger \Phi_u + c_9 \Phi^\dagger_d \Phi_d.
\end{eqnarray}

We have checked that the total Higgs potential resulting from
above admits the required forms for the VEVs as extrema for the case where all the Higgs
potential parameters are real. Terms from $a_{1,2}$,
$b_{1,2}$ and $c_1$ allow two solutions for the VEV pattern of $\chi$,
namely, that all component VEVs are equal or that only one of the them is nonzero.
The latter is the desired one. Terms from
$a_{5,6,7,8}$, $b_{5,6,7,8}$ and $c_{7,8,9}$ force the component
VEVs to be equal, if nonzero, in each of the fields $\chi'$,
$\Phi_{u}$ and $\Phi_d$.

\end{document}